\documentclass[twocolumn]{aastex631}

\hypersetup{linkcolor=blue, citecolor=blue, filecolor=cyan, urlcolor=blue}

\usepackage{multirow}
\usepackage{amsmath}

\shorttitle{Morphology and size of SMGs}
\shortauthors{J. Ren et al.}
\graphicspath{{./}{figures/}}
\begin{document}

\correspondingauthor{F. S. Liu and Nan Li}
\email{E-mail: fsliu@nao.cas.cn; nan.li@nao.cas.cn}

\title{The Evolution of Size and Merger Fraction of Submillimeter Galaxies across $1 < z \lesssim 6$ as Observed by JWST}

\author[0000-0002-5043-2886] {Jian Ren}
\affil{National Astronomical Observatories, Chinese Academy of Sciences, 20A Datun Road, Chaoyang District, Beijing 100101, China}
\affil{Key Laboratory of Space Astronomy and Technology, National Astronomical Observatories, Chinese Academy of Sciences, 20A Datun Road, Chaoyang District, Beijing 100101, China} 

\author[0009-0001-7105-2284]{F. S. Liu $^{\color{blue} \dagger}$}
\affil{National Astronomical Observatories, Chinese Academy of Sciences, 20A Datun Road, Chaoyang District, Beijing 100101, China}
\affil{Key Laboratory of Optical Astronomy, National Astronomical Observatories, Chinese Academy of Sciences, 20A Datun Road, Chaoyang District, Beijing 100101, China}
\affil{School of Astronomy and Space Science, University of Chinese Academy of Science, Beĳing 100049, China}

\author{Nan Li $^{\color{blue} \dagger}$}
\affil{National Astronomical Observatories, Chinese Academy of Sciences, 20A Datun Road, Chaoyang District, Beijing 100101, China}
\affil{Key Laboratory of Space Astronomy and Technology, National Astronomical Observatories, Chinese Academy of Sciences, 20A Datun Road, Chaoyang District, Beijing 100101, China}
\affil{School of Astronomy and Space Science, University of Chinese Academy of Science, Beĳing 100049, China}

\author{Pinsong Zhao}
\affil{Kavli Institute for Astronomy and Astrophysics, Peking University, Beijing 100871, China}
\affil{National Astronomical Observatories, Chinese Academy of Sciences, 20A Datun Road, Chaoyang District, Beijing 100101, China}

\author[0009-0001-5320-1450]{Qifan Cui}
\affil{Shanghai Key Lab for Astrophysics, Shanghai Normal University, Shanghai 200234, China}
\affil{National Astronomical Observatories, Chinese Academy of Sciences, 20A Datun Road, Chaoyang District, Beijing 100101, China}

\author{Qi Song}
\affil{National Astronomical Observatories, Chinese Academy of Sciences, 20A Datun Road, Chaoyang District, Beijing 100101, China}
\affil{Key Laboratory of Space Astronomy and Technology, National Astronomical Observatories, Chinese Academy of Sciences, 20A Datun Road, Chaoyang District, Beijing 100101, China}

\author{Yubin Li}
\affil{National Astronomical Observatories, Chinese Academy of Sciences, 20A Datun Road, Chaoyang District, Beijing 100101, China}
\affil{Key Laboratory of Space Astronomy and Technology, National Astronomical Observatories, Chinese Academy of Sciences, 20A Datun Road, Chaoyang District, Beijing 100101, China}

\author{Hao Mo}
\affil{National Astronomical Observatories, Chinese Academy of Sciences, 20A Datun Road, Chaoyang District, Beijing 100101, China}
\affil{Key Laboratory of Optical Astronomy, National Astronomical Observatories, Chinese Academy of Sciences, 20A Datun Road, Chaoyang District, Beijing 100101, China}
\affil{School of Astronomy and Space Science, University of Chinese Academy of Science, Beĳing 100049, China}

\author{Hassen M. Yesuf}
\affil{Key Laboratory for Research in Galaxies and Cosmology, Shanghai Astronomical Observatory, Chinese Academy of Sciences, 80 Nandan Road, Shanghai 200030, China}

\author{Weichen Wang}
\affil{Dipartimento di Fisica G. Occhialini, Università degli Studi di Milano-Bicocca, Piazza della Scienza 3, I-20126 Milano, Italy}

\author[0000-0001-7943-0166]{Fangxia An}
\affil{Purple Mountain Observatory, Chinese Academy of Sciences, 10 Yuanhua Road, Nanjing 210034, China}

\author{Xian Zhong Zheng}
\affil{Tsung-Dao Lee Institute and Key Laboratory for Particle Physics, Astrophysics and Cosmology, Ministry of Education, Shanghai Jiao Tong University, Shanghai 200240, China}
\affil{Purple Mountain Observatory, Chinese Academy of Sciences, 10 Yuanhua Road, Nanjing 210034, China}

\begin{abstract}

Precise tracking of the growth in galaxy size and the evolution of merger fractions with redshift is vital for understanding the formation history of submillimeter galaxies (SMGs). This study investigates these evolutions over a broad redshift range ($1 < z \lesssim 6$), using a sample of 222 SMGs with a median redshift of $z = 2.61^{+0.89}_{-0.82}$ identified by ALMA and JCMT, enhanced by the advanced imaging capabilities of the JWST/NIRCam and MIRI. We find significant evolution in effective radii ($R_e$) in rest-frame V-band ($R_e \propto (1 + z)^{-0.87 \pm 0.08}$) and near-infrared (NIR) band ($R_e \propto (1 + z)^{-0.88 \pm 0.11}$), with the NIR size evolution resembling that of massive star-forming galaxies at lower redshift. Visual inspections reveal a major merger fraction of $24.3 \pm 3.7\%$ and an interaction fraction of up to $48.4 \pm 11.1\%$. The major merger fraction exhibits an increase from 14.7$\pm9.1$\% at $z = 1$ to 26.6$\pm 8.4$\% at $z = 3$, after which it remains approximately constant across the redshift range $3 < z < 6$. In contrast, the interaction fraction remains relatively stable across the range $2 < z < 5$.  Our results indicate that late-stage major mergers are not the primary formation mechanism for SMGs at $z<3$, while interactions appear to play a significant role across the broader redshift range of $1<z<6$. Additionally, HST-based major merger identifications may overestimate the true fraction by a factor of 1.7 at $z \sim 2$. These findings highlight the varying roles of mergers and interactions in driving the formation of massive, dusty star-forming galaxies across different redshifts.

\end{abstract}


\section{Introduction} \label{sec:intro}

Submillimeter galaxies (SMGs) represent a unique class of astronomical objects that have been extensively observed at submillimeter and far-infrared wavelengths. They are characterized by high gas content  \citep{Tacconi2006, Riechers2010}, high star formation rates (SFRs) \citep{Barger2014}, and active galactic nuclei (AGN) activity \citep{Ueda2018}. These galaxies indicate intense dusty starburst activity and significant mass assembly. Typically detected at high redshifts, SMGs are considered the high-redshift counterparts of local ultra-luminous infrared galaxies (ULIRGs). They are ideal laboratories for investigating obscured star formation processes and chemical enrichment in the early universe.

Previous research has indicated a significant number of major merger events in SMGs at cosmic noon ($z \sim$2). High-resolution imaging observations from the Hubble Space Telescope (HST) Advanced Camera for Surveys (ACS) suggest that the merger fraction exceeds $60\%$ \citep{Chapman2003, Conselice2003, Smail2004}. Observations from the HST Wide Field Camera 3 (WFC3) F160W imaging indicate that the merger fraction in SMGs or ULIRGs is approximately $50\%$ or higher \citep{ Kartaltepe2012, Chen2015, Stach2019}.  However, some studies have pointed out significant morphological differences in SMGs across rest-frame ultraviolet (UV) and optical wavelengths, suggesting that irregular morphologies may be more heavily influenced by dust obscuration rather than mergers \citep{Swinbank2010}.  Additionally, numerical simulation studies have yielded divergent results. Some research suggests that a considerable fraction of SMGs are driven by intense starbursts triggered by mergers \citep{Baugh2005, Narayanan2010, Hayward2013}. In contrast, other studies, based on hydrodynamical simulations, propose that the observed properties of SMGs can be attributed to non-merging massive galaxies that maintain high star formation rates for approximately 1 billion years without deviating from the main sequence \citep{Narayanan2015}. This has raised questions about whether mergers are the primary driver of star formation in SMGs. 

Due to the effects of dust extinction and star formation activity, the abundance of dust significantly influences the morphology of SMGs in the rest-frame optical band \citep{Smail1999}. In the early universe (i.e., at redshifts $z>4$), these galaxies are almost invisible in HST observations \citep{Smail1998, Enia2022}. In contrast, the rest-frame near-infrared (NIR) band, less affected by dust extinction, predominantly reflects the radiation from old stellar populations.  Studying SMGs' rest-frame NIR morphology and size can provide a clearer understanding of their stellar mass distribution, assembly processes, and formation scenarios.

With the release of James Webb Space Telescope (JWST) Near-Infrared Camera (NIRCam) data, further research on the properties of SMGs during the cosmic noon era ($1 < z < 3$) has emerged \citep{ Cheng2022, Cheng2023, Boogaard2023, Gillman2023, Huang2023, McKinney2024}. Notably, \citet{Cheng2023} studied 16 faint SMGs using the Atacama Large Millimeter/submillimeter Array (ALMA) and JWST/NIRCam data, revealing a diverse range of properties including both star-forming and quiescent galaxies.  Most SMGs in their sample are compact disk galaxies, with a median half-mass radius of 1.6\,kpc, showing little to no signs of disturbance. This suggests that secular growth could be a potential pathway for the assembly of high-mass disk galaxies. Additionally, JWST/MIRI data have been utilized to investigate the size and morphology of several high-redshift ($z\sim4$) extreme dusty star-forming galaxies (DSFGs) \citep{Smail2023, Gomez2024}. 

 Recently, \citet{McKinney2024} conducted an in-depth analysis of 289 sub-millimeter galaxies (SMGs) in the COSMOS-Web field using the latest multi-band imaging data from the JWST.  Their study revealed critical parameters such as galaxy mass, star-formation rate, and optical extinction. Notably, they found that approximately 30\% of these galaxies had no previous optical/near-infrared detection records. Furthermore, their research indicated that about 17\% of SMGs are undergoing major mergers. \citet{Gillman2024} utilized JWST/NIRCam and MIRI data to study the morphological and structural properties of 80 SMGs detected by ALMA within the redshift range of $1 < z < \sim 5$. The study found that approximately 60\% of the SMGs appeared to be in late-stage major mergers or potential minor mergers, while about 40\% displayed undisturbed disk-like morphologies. Additionally, the authors demonstrated that SMGs have comparable near-infrared sizes to less active galaxies but exhibit lower S\'ersic indices. This suggests the presence of bulge-less disks with significant structured dust content, leading to distinct morphological characteristics. Despite these advances, there is still controversy regarding whether galaxy mergers play a decisive role in the evolution of SMGs at different redshifts and whether the sizes of SMGs evolve with redshift.

In this work, we construct a comprehensive sample of 222 SMGs with redshifts ranging from $1 < z < 6$ and dust luminosity $\log(L_{\rm IR}/L_{\odot}) > 12$. These SMGs were observed by the ALMA and the James Clerk Maxwell Telescope (JCMT) within the COSMOS and UDS fields. Utilizing the latest JWST/NIRCam and MIRI data, we conduct a detailed study of the effective radii in the rest-frame V-band and near-infrared (NIR) band of these SMGs, along with their merger fraction. This study traces their evolution from the cosmic dawn to the cosmic noon. 

The paper is structured as follows: Section 2 describes the JWST/NIRCam and MIRI imaging data. Section 3 details the sample selection and the properties obtained, leading to the main results and discussion in Sections 4 and 5. A brief summary is provided in Section 6. Throughout this study, we adopt a concordance cosmology with $H_0 = 70 \, \text{km s}^{-1} \text{Mpc}^{-1}$, $\Omega_{\rm m} = 0.3$, and $\Omega_{\Lambda} = 0.7$.

\section{Data} \label{sec:data}

\subsection{JWST imaging data}

We acquired raw broad-band JWST NIRCam imaging data for the COSMOS field, encompassing filters F090W to F444W, through programs 1727 \citep{Pro1727}, 1810 \citep{Pro1810}, 1837 \citep{Pro1837}, 1840 \citep{Pro1840}, and 2514 (PI, Christina Williams). Similar data for the UDS field were obtained from programs 1837, 2514, and 3990 \citep{Pro3990}. The raw MIRI imaging data in the COSMOS field, captured through filters F770W and F1800W, were sourced from programs 1727 and 1837. The raw MIRI data in the UDS field was derived from program 1837. Our analysis utilized JWST data collected in the COSMOS field before January 7, 2024, and data collected in the UDS field before January 21, 2024.

Our data reduction process began with uncalibrated data, applying Stage 1 of the JWST pipeline with standard parameters for initial detector corrections. We then addressed ``snowball" artifacts, followed by Stage 2, where we subtracted ``wisp" artifacts and reduced $1/f$ noise using the method described by \citet{Schlawin2020}. Masks were applied to correct for persistence, dragon breath, ginkgo leaf, wisp features, and other artifacts. For Stage 3, we performed a custom astrometric correction and executed the \texttt{OutlierDetection} step. This was followed by a tailored outlier detection process and background subtraction before resampling. We initiated the World Coordinate System (WCS) calibration on JWST/F150W images, utilizing HST/F160W images from the CANDELS survey \citep{Grogin2011, Koekemoer2011, Faber2011}. Subsequently, we extended this calibration process to other JWST bands, leveraging the JWST/F150W images as a reference. Mosaics were constructed for each observation, with a custom background subtraction applied to sub-mosaics to refine the final product.

The MIRI data underwent a similar pipeline process, incorporating super-sky flats in Stage 2 along with custom corrections for stripe and background effects. Stage 3 for MIRI followed a procedure analogous to that of NIRCam, with the primary difference being the reference catalog used for WCS alignment, which was sourced from F444W images.

The reduced NIRCam data covered 445 arcmin$^2$ in the COSMOS field and 273 arcmin$^2$ in the UDS field at F444W. The MIRI data spanned 259 arcmin$^2$ in the COSMOS field and 129 arcmin$^2$ in the UDS field at F770W, and all drizzled to a pixel scale of $0.^{\prime \prime}03$. The empirical full width at half maximum (FWHM) of the point spread function (PSF) is $0.^{\prime \prime}165$ at F444W and $0.^{\prime \prime}293$ at F770W. 
This work is part of the Spatially Pixel-level Resolved Investigations into Nascent Galaxies with the James Webb Space Telescope (JWST-SPRING) project. This program aims to provide the astronomical community with extensive, homogeneous, and deep imaging data from JWST's NIRCam+MIRI, as well as spectroscopic data from NIRSpec and NIRCam/WFSS, all precisely matched with observations from the Hubble Space Telescope (HST). The JWST-SPRING dataset will encompass all previous deep extragalactic survey observations by HST, covering the five CANDELS fields -- COSMOS, EGS, GOODS-North, GOODS-South, and UDS -- as well as the broader COSMOS field and numerous smaller regions. For more information about the JWST-SPRING data, please visit the project website\footnote{\url{http://groups.bao.ac.cn/jwst_spring/}} and refer to the overview paper (Liu et al., in preparation).
%

\subsection{Submillimeter sources}

The SCUBA-2 COSMOS survey (S2COSMOS), conducted between 2016 and 2017, utilized the SCUBA-2 instrument at the JCMT to map the COSMOS field at 850\,$\mu$m. As reported by \citet{Simpson2019}, the survey delineated a 1.6 deg$^2$ MAIN region with a median noise level of 1.2 mJy beam$^{-1}$ and a 1 deg$^2$ SUPP region with a sensitivity of 1.7 mJy beam$^{-1}$. The survey detected 1,020 and 127 submillimeter sources in the MAIN and SUPP regions, respectively, with a false detection rate of 2\%, employing a PSF with a full width at half maximum (FWHM) of $14.\!\!^{\prime\prime}8$.

The Automated Mining of the ALMA Archive in the COSMOS Field ($\rm A^3COSMOS$) project, led by \citet{Liu2019a}, developed automated mining pipelines for the ALMA archive in the COSMOS field to enhance the usability of these data. By applying stringent selection criteria to ALMA detections, the project aimed to reduce the incidence of spurious sources, resulting in a catalog of $\sim$ 700 galaxies within the redshift range $0.3 < z < 6$, with a minimal spurious fraction. The latest $\rm A^3COSMOS$ database comprises around 4,000 pipeline-processed continuum images from the public ALMA archive, yielding 2,050 unique detected sources, including those with and without a known optical counterpart \citep{Adscheid2024}.

The AS2UDS catalog encompasses 716 single-dish submillimeter sources identified in the UKIDSS/UDS field through an 870\,$\mu$m continuum survey with ALMA \citep{Stach2019}. The properties of these SMGs have been studied by \citet{Dudzeviciute2020}.

\section{Sample selection and analyses} \label{sec:data}

\begin{figure}[t!]
\centering
\includegraphics[width=1.0\columnwidth] {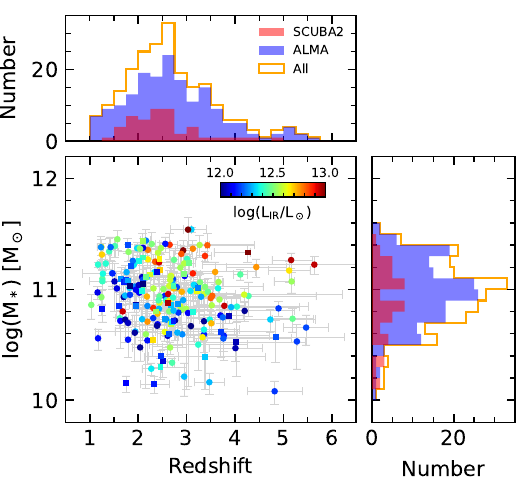} 
\caption{The stellar mass and redshift for the 222 SMGs. Data points are color-coded by dust luminosity. Squares represent JCMT-detected sources, while circles represent ALMA-detected sources. The top panel displays the redshift distribution, and the right panel shows the stellar mass distribution.}
\label{fig:Fig01}
\end{figure}

\subsection{Sample Selection and Photometry}

We collected the S2COSMOS and $\rm A^3COSMOS$ catalogs from the COSMOS field and the AS2UDS catalog from the UDS field. We then cross-matched all SMGs with sources observed by the JWST NIRCam and MIRI. For ALMA-observed sources, the resolution was sufficient to identify optical counterparts. Conversely, for JCMT/SCUBA-2 observed sources in the COSMOS field, we relied on the optical/near-infrared/radio counterparts identified by \citet{An2019} using a machine learning-assisted method \citep{An2018}. In total, we identified 453 submillimeter sources with observations in at least one JWST filter from both the COSMOS and UDS fields.

The integrated fluxes of SMGs in the HST bands (F814W, F105W, F125W, F140W, F160W) and JWST bands (F090W, F115W, F150W, F200W, F277W, F356W, F444W, F770W) were measured by accumulating the radial surface brightness profiles. These profiles were computed using the Python package \texttt{photutils} \citet{Bradley2024} on PSF-matched images with all contaminations masked. We also cross-matched our sample with the COSMOS2015 catalog \citep{Laigle2016} in the COSMOS field and multi-wavelength photometry data of SMGs in the UDS field \citep{Michalowski2017, Dudzeviciute2020} to obtain additional far-IR data.

 \subsection{Redshifts}
Initially, we utilized the \texttt{CIGALE} SED fitting code \citep{Boquien2019, Yang2020, Yang2022} to estimate the photometric redshifts ($z_{\rm phot}$) of our SMG sample. Subsequently, we cross-matched our SMG sample in the COSMOS field with the publicly available spectroscopic redshift ($z_{\rm spec}$) and grism redshift ($z_{\rm grism}$) data of the COSMOS field \citep{Kriek2015, Momcheva2016, Hasinger2018}. Using the matched $z_{\rm spec}$, we assessed the precision of the $z_{\rm phot}$ from the \texttt{CIGALE} sample with more than five bands, resulting in a precision of $\lvert \Delta z \rvert /(1+z_{\rm spec}) = 0.09$. This precision is comparable to the $\lvert \Delta z \rvert /(1+z_{\rm spec}) = 0.12$ obtained by \citet{Uematsu2024} and $\sim$ 0.1 obtained by \citet{McKinney2024} using  \texttt{CIGALE} for SED fitting of SMGs in the COSMOS field.

Concurrently, we cross-matched the SMG samples with the COSMOS-ReGEM catalog, in which the photoz is the median value of the COSMOS2020 CLASSIC catalog \citep{Weaver2022}, the COSMOS Photometric Redshifts with 30-Bands catalog \citep{Ilbert2009}, the UrtaVISTA catalog \citep{Muzzin2013}, and the COSMOS2015 catalog \citep{Laigle2016} in the COSMOS field. A detailed description can be found in \citet{Ren2023}. For our SMG sample, the ReGEM catalog provides a $z_{\rm phot}$ precision of approximately $\lvert \Delta z \rvert /(1+z_{\rm spec}) = 0.06$. Ultimately, in the COSMOS field, the priority for adopting the SMG best redshift ($z_{\rm best}$) is as follows: $z_{\rm spec}$ first, then the $z_{\rm phot}$ from ReGEM, and lastly the $z_{\rm phot}$ estimated by our \texttt{CIGALE}. Consequently, the $z_{\rm best}$ precision in the COSMOS field part is significantly better than $\lvert \Delta z \rvert /(1+z_{\rm spec}) = 0.09$.

For sources in the UDS field, we first used the best redshifts provided by the CANDELS catalog \citep{Skelton2014, Momcheva2016}, which mainly include $z_{\rm spec}$, $z_{\rm grism}$, and $z_{\rm phot}$. Then we considered the median of the three measurements: the $z_{\rm phot}$ with and without IR from \citet{Dudzeviciute2020}, and our \texttt{CIGALE} $z_{\rm phot}$. Finally, we used our \texttt{CIGALE} $z_{\rm phot}$. We note that four SMGs in our sample in the UDS field were given JWST/NIRSpec spectra in \citet{Cooper2024}, and we adopted the $z_{\rm spec}$ for these four SMGs.

\begin{table*}[]
\centering
\scriptsize
\caption{Detection Counts of SMGs in HST and JWST Bands.}
\label{Table1}
\begin{tabular}{|c|c|c|c|c|c|c|c|c|c|c|c|c|c|c|}
\hline
 Field  & F814W & F090W & F105W & F115W & F125W & F140W & F150W & F160W & F200W & F277W & F356W & F444W & F770W  &  Total \\
\hline
COS   & 117      & 65         & 18         & 156      &  90        & 62        & 154       & 103       & 75        & 162       & 74         & 154        & 99      & 163 \\
\hline 
UDS   &  34       & 53         & 5           & 56        & 35         & 24        & 57         & 36         & 56        & 59         & 59         & 59         & 26        &  59 \\
\hline
All    &  151     & 118        & 23          & 212      & 125       & 86        & 211       & 139      & 131       & 221      & 133       & 213        &125      &  222\\
\hline
\end{tabular}
 \end{table*}

\subsection{SED fitting}

 In our analysis, we employed \texttt{CIGALE} to conduct SED fitting for our SMG sample, with photometric data spanning from optical to submillimeter wavelengths. During the SED-fitting process, we utilized the \texttt{sfhdelayed} module. This module incorporated a main stellar population with ages ranging from 500\,Myr to the age of the universe at the redshift of each SMG, and an e-folding time spanning from 100 to 10000\,Myr. We adopted the \texttt{BC03} stellar population module \citep{Bruzual2003}, which was combined with a Chabrier initial mass function \citep[IMF;][]{Chabrier2003}. Dust attenuation was modeled using the \texttt{dust-modified-starburst} model \citep{Calzetti2000}, with $E(B-V)_{\rm lines}$ values ranging from 0.1 to 3. A factor of 0.44 was applied to $E(B-V)$ to correct for differential dust attenuation between old and young stellar populations \citep{Calzetti2000, Charlot2000, Wild2011}. Dust emission was accounted for using the \texttt{Dale2014} prescription \citep{Dale2014}. Additionally, the \texttt{fritz2006} AGN module \citep{Fritz2006} was incorporated into our SED fitting, allowing for an AGN fraction ranging from 0 to 0.5. This enabled us to account for the potential contribution of AGN activity to the overall SED. For additional details on the use of other modules and parameter settings in fitting SMG properties using \texttt{CIGALE}, one can refer to \citet{McKinney2024, Uematsu2024, Uematsu2025}.

Furthermore, a mock analysis was included in the fitting process to assess the reliability of our results. The dispersion between the SED results and the mock results was 0.09\,dex for stellar mass ($M_*$) and 0.11\,dex for $L_{\rm IR}$, with no systematic bias observed. This indicates that our SED fitting method is robust and provides accurate estimates of the SMG properties.

\subsection{Final Sample}

Our final selection of SMGs is derived from a larger pool of 453 SMGs observed by JWST across two fields and is a carefully curated subset. The curation process is based on three principal criteria. First, we excluded sources with limited HST and JWST band coverage, as such constraints impede the precise determination of physical properties via SED fitting. Second, to define our sample selection boundary, we omitted sources with redshifts $z<1$. Lastly, we filtered out sources that did not meet our thresholds for stellar mass and total infrared luminosity. These refinements have significantly enhanced the representativeness of our sample in terms of mass and infrared luminosity completeness. Our refined sample comprises 222 SMGs with stellar masses greater than $\log(M_*/M_\odot) > 10$ and total infrared luminosities satisfying $12 < \log(L_{\rm IR}/L_\odot) < 13$, across the redshift range of $1 < z < 6$. In our sample, 46 SMGs overlap with those in \citet{Gillman2024}.

 Within this sample, 76 (34\%) SMGs have spectroscopic or grism redshifts. The median stellar mass of our sample is $\log(M/M_\odot) = 10.97^{+0.29}_{-0.33}$, the median redshift is $z = 2.61^{+0.89}_{-0.82}$, and the median infrared luminosity is $\log(L_{\rm IR}/L_\odot) = 12.38^{+0.28}_{-0.24}$. The distribution of stellar masses and redshifts, along with the infrared luminosities of our sample, is depicted in Figure \ref{fig:Fig01}. Table \ref{Table1} shows the number of our final samples detected in the JWST and HST filters. The mean filters used for SED fitting of our SMGs are 12 filters spanning from the optical to ALMA band 3. In our sample, the detections from mid-infrared to far-infrared are as follows: 99 (45\%) at MIPS 24\,$\mu$m, 22 (10\%) at 100\,$\mu$m, 30 (14\%) at 160\,$\mu$m, 88 (40\%) at 250\,$\mu$m, 78 (35\%) at 350\,$\mu$m, and 42 (19\%) at 500\,$\mu$m.

We present a comparison of the SMG properties derived from our \texttt{CIGALE}-based SED fitting with those from previously published SMG catalogs. The stellar masses derived from \texttt{MAGPHYS} \citep{DaCunha2008} are larger than our results by 0.1-0.2\,dex in the $\rm A{^3}COSMOS$ \citep{Adscheid2024} and the AS2UDS catalog \citep{Dudzeviciute2020}. \citet{Uematsu2024} found that the stellar mass derived from \texttt{CIGALE} is 0.3-0.4\,dex lower than that derived from \texttt{MAGPHYS} due to the different dust attenuation models adopted in SED fitting. A similar difference is observed in the H-band mass-to-light ratio recovered by \texttt{CIGALE}, which is $\sim$0.2\,dex less than the average among ALESS SMGs derived from \texttt{MAGPHYS} in \citet{McKinney2024}. Considering a deviation of $\sim$ 0.2\,dex, our median stellar mass is consistent with the median stellar masses provided by $\rm A{^3}COSMOS$ \citep{Adscheid2024}, SCUBADive \citep{McKinney2024}, and AS2UDS \citep{Dudzeviciute2020}. The $L_{\rm IR}$ shows systematic deviations of 0.04\,dex between our results and $\rm A{^3}COSMOS$ \citep{Adscheid2024} and AS2UDS \citep{Dudzeviciute2020}. Since our stellar mass is primarily used to select SMGs with $\log(M_*/M_\odot) > 10$, and SMGs with $\log(M_*/M_\odot) < 10.5$ actually make up a very small fraction of our sample, as shown in Figure \ref{fig:Fig01}, even considering an underestimation of $\sim$ 0.2\,dex in stellar mass compared to previous work, a new selection criterion of $\log(M_*/M_\odot) > 9.8$ would not significantly affect our sample.

Among our 222 SMGs, 163 are in the COSMOS field, with 48 from the S2COSMOS survey \citep{Simpson2019} based on JCMT/SCUBA-2 observations and the remaining 115 from the $\rm A{^3}COSMOS$ catalog \citep{Liu2019a,Adscheid2024} based on ALMA observations. In the UDS field, there are 59 SMGs, all from the AS2UDS survey \citep{Stach2019} based on ALMA observations. The optical and near-infrared counterparts of the 48 S2COSMOS sources were identified by \citet{An2019} using a machine learning method with an accuracy exceeding 85\%. This implies that fewer than 8 uncertain samples exist among these 48 sources. The remaining sources, which are based on ALMA observations, are more reliable. Therefore, the uncertainty of SMGs in our final sample is considerably less than 4\%.

\begin{figure*}[!ht]
\centering
\includegraphics[width=1.0\textwidth] {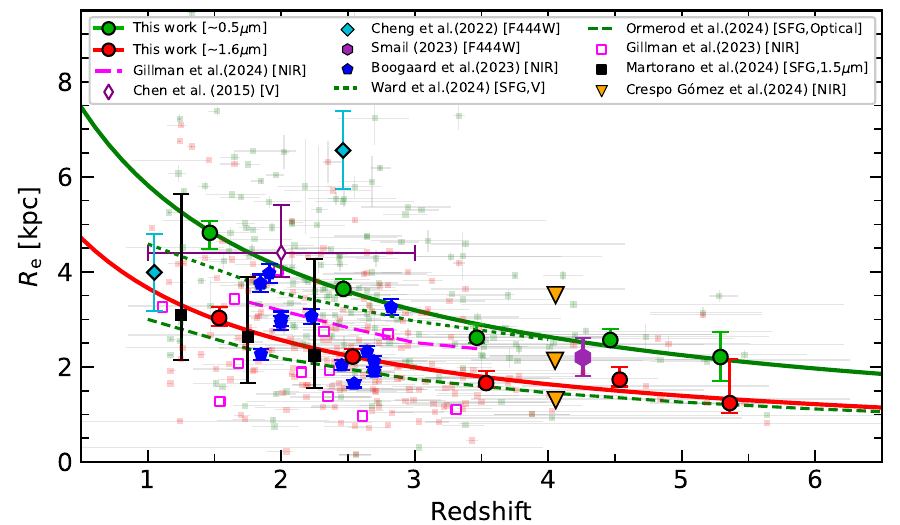} 

\caption{
{\bf Effective Radius as a Function of Redshift for Our SMGs.} The green circles denote the median $R_e$ of our SMGs in the rest-frame V-band,  while the red circles represent the median $R_e$ of our SMGs in the rest-frame NIR-band for each redshift bin. The X-axis positions of the red and green circles are shifted by $\pm$0.03 for clarity. The solid green and red lines represent the best-fitting $R_e$ evolution with redshift for our sample in the rest-frame V-band and NIR-band, respectively.  The error bars on our green and red data points represent the uncertainties of the median value in each redshift bin. The magenta dashed line traces the NIR ($\sim$ 1$\mu$m) size evolution of luminous ALMA-detected SMGs in the COSMOS and UDS fields, as measured with JWST/NIRCam and MIRI by \citet{Gillman2024}. The green dotted line delineates the rest-frame optical size evolution of star-forming galaxies for log($M_*/M_\odot$)$>10.5$ at $0.5<z<5$, while the green dashed line shows the size evolution for log($M_*/M_\odot$)$>9.5$ at $0.5<z<8$, both from the JWST/CEERS survey by \citet{Ward2024} and \citet{Ormerod2024}, respectively. Cyan markers represent SMGs in the SMACS J0723.3–7327 field, with sizes derived from JWST/F444W by \citet{Cheng2022}. Unfilled magenta squares represent SCUBA-2 SMGs selected within the JWST/CEERS survey by \citet{Gillman2023}. Purple hexagons highlight an SMG with log($M_*/M_\odot$)$\sim$11.8 at $z=4.26$ in the Prime Extragalactic Areas for Reionization and Lensing Science (PEARLS) field, with size derived from JWST/NIRcam F444W by \citet{Smail2023}. Blue pentagons correspond to SMGs with log($M_*/M_\odot$)$>10$ and redshifts between $1.8<z<4.6$ in the Hubble Ultra Deep Field (HUDF), with sizes determined from JWST/MIRI F560W by \citet{Boogaard2023}. Orange triangles represent three SMGs with very high IR luminosities ($L_{\rm IR}>10^{13}$ L$_\odot$), with sizes determined from JWST/MIRI F560W by \citet{Gomez2024}. The unfilled purple diamonds represent the rest-frame V size at $z\sim$2 given by \citet{Chen2015} using HST/F160W data. The black squares indicate the median size of massive star-forming galaxies ($11<\log(M_*/M_\odot)<11.5$) at rest-frame 1.5\,$\mu$m, measured by \citet{Martorano2024} for the redshift range of $0.5<z<2.5$.
}

\label{fig:Fig02}
\end{figure*}

\subsection{Size measurements}

Our objective is to examine the size evolution of these SMGs separately in the rest-frame V-band and rest-frame NIR-band. To accomplish this, we selected the rest-frame V and NIR-bands using the rest-frame wavelengths corresponding to the F115W to F770W bands at different redshifts. Specifically, we chose the JWST band closest to 0.5\,$\mu$m within the 0.4--0.6\,$\mu$m range to represent the rest-frame V-band, and the band closest to 1.6\,$\mu$m (H-band) within the 1--2\,$\mu$m range to represent the rest-frame NIR-band. For example, for an SMG at $z = 1.85$, the rest-frame 0.4--0.6\,$\mu$m includes the F115W and F150W bands, but the rest-frame wavelength of F150W is closest to 0.5\,$\mu$m. Meanwhile, the rest-frame 1--2\,$\mu$m range encompasses the F356W and F444W bands, with the rest-frame wavelength of F444W being closer to 1.6\,$\mu$m. Consequently, for this SMG, the rest-frame V-band is F150W, and the rest-frame NIR-band is F444W.

We utilized four separate S\'{e}rsic fitting packages—\texttt{PetroFit} \citep{Geda2022}, \texttt{Pysersic} \citep{Pasha2023}, \texttt{GALFIT} \citep{Peng2002}, and \texttt{Statmorph} \citep{Rodriguez-Gomez2019}—to perform independent S\'{e}rsic fittings. As a result, the S\'{e}rsic index ($n$) and effective radius ($R_e$) for the rest-frame V and NIR-band of each SMG were measured using these four tools on the contaminations-masked images. By comparing the fitting results from different software, we found that the systematic bias in the fitted $R_{\rm e}$ values obtained from different software is as small as $0.^{\prime \prime}01$ in the rest-frame V-band and $0.^{\prime \prime}004$ in the rest-frame NIR-band. Among them, the fitting results from \texttt{PetroFit} exhibit fewer outliers compared to other software. Therefore, if the fitting result from \texttt{PetroFit} lies between the results of the other three softwares, we adopt the \texttt{PetroFit} result. If \texttt{PetroFit} does not provide a fitting value or its result falls outside the range of the other three software results, we then use the median value from the remaining three softwares. This approach effectively reduces the impact of outliers that can arise from relying on a single fitting software.

There are 179 (80\%) sources with measured $R_e$ in the V-band, with a median $R_e$ and 16th to 84th percentile intervals of $3.60^{+2.19}_{-1.44}$\,kpc. In the NIR-band, there are 186 (81\%) sources with measured $R_e$, with a median $R_e$ and 16th to 84th percentile intervals of $2.26^{+1.53}_{-0.99}$\,kpc. There are 158 SMGs (68\%) having both V-band and NIR-band size measurements.

\section{RESULTS}

\begin{figure*}[ht!]
\centering
\includegraphics[width=0.99\textwidth] {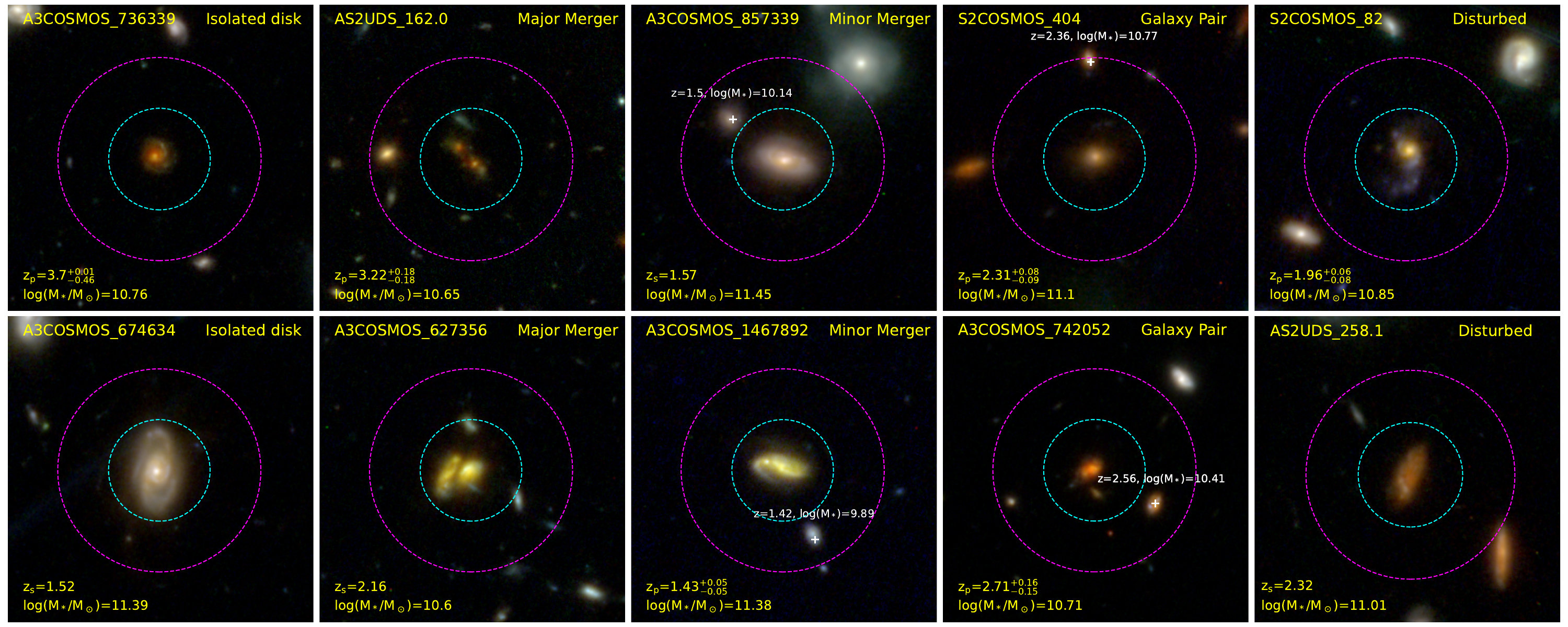} 
\caption{{\bf Example images of SMGs classified by visual identification}. The radius of cyan and magenta circle are 15\,kpc and 30\,kpc respectively.}
\label{fig:Fig03}
\end{figure*}

\subsection{Size evolution}
 
\begin{table}[t!]
\centering
\scriptsize
\caption{ Number of SMGs used in the size evolution fitting.}
\label{Table2}
\begin{tabular}{|c|c|c|c|c|c|}
\hline 
 Range   & $R_{\rm e}$ (V)  & N$_{\rm V}$& $R_{\rm e}$ (NIR)  & N$_{\rm NIR}$  \\
 {}   &  (kpc)            & {}                  &          (kpc)     & {}    \\
\hline

$1<z<2$   & 4.82$^{+1.72}_{-2.27}$    & 47 (92\%)   & 2.93$^{+1.71}_{-0.99}$    &46 (90\%)   \\

\hline

$2<z<3$   & 3.64$^{+1.84}_{-1.18}$    & 83 (80\%)   & 2.22$^{+1.51}_{-0.76}$    &93 (89\%)  \\

\hline
$3<z<4$   & 2.61$^{+1.71}_{-0.63}$    & 33 (70\%)   & 1.66$^{+1.48}_{-0.54}$    &35 (74\%)   \\

\hline

$4<z<5$   & 2.57$^{+0.75}_{-0.42}$    & 10 (77\%)   & 1.73$^{+0.74}_{-0.41}$    &8 (62\%)      \\

\hline

$5<z<6$   & 2.21$^{+1.29}_{-1.22}$    & 6 (86\%)     & 1.24$^{+1.86}_{-0.43}$   &4 (57\%)    \\

\hline

\end{tabular}
 \end{table}

We performed a detailed analysis to examine the relationship between the effective radius ($R_e$) of SMGs in both the rest-frame V-band ($\sim$ 0.5\,$\mu$m) and rest-frame NIR-band ($\sim$1.6\,$\mu$m) and redshift, as shown in Figure \ref{fig:Fig02}. The median size, as well as the 16th to 84th percentile intervals, the number, and fraction with size measurements in each redshift bin are provided in Table \ref{Table2}. Our results indicate that in the V-band, the average effective radius ($R_e$) of SMGs significantly increases, from 2\,kpc at $z=6$ to 5.5\,kpc at $z=1$. A similar trend is observed in the NIR band, where $R_e$ increases from $\sim$1.3\,kpc to $\sim$3.5\,kpc within the same redshift interval. This suggests that the spatial distribution of star formation activity within SMGs is more extended at lower redshifts, indicating that the stellar mass may be more centrally concentrated than in unobscured star-forming regions.

 By fitting the $R_e - z$ relationship, we find that the power-law function exhibits smaller AIC and BIC values, as well as a smaller chi-square value, compared to other simple functional forms such as linear functions. This suggests that the power-law model is better suited to describe the evolution of SMG size with redshift.  The derived $R_e - z$ relations are as follows: in the V-band, $R_e = (10.62 \pm 1.03) \times (1+z)^{-0.87 \pm 0.08}$; in the NIR-band, $R_e = (6.74 \pm 0.93) \times (1+z)^{-0.88 \pm 0.11}$. These relations provide a quantitative framework for understanding the cosmic evolution of sizes in SMGs.

Additionally, we find that the rest-frame NIR sizes of SMGs in previous studies based on JWST observations generally align with the evolutionary trend we observed at $z<4$. However, several recent studies \citep{Smail2023, Gomez2024, Gillman2024} indicate that the average sizes of massive SMGs are larger than those of the SMGs we studied. This discrepancy may be due to differences in sample selection and rest-frame band selection.

\subsection{Merger fraction evolution}

\begin{figure*}[!ht]
\centering
\includegraphics[width=1.0\textwidth] {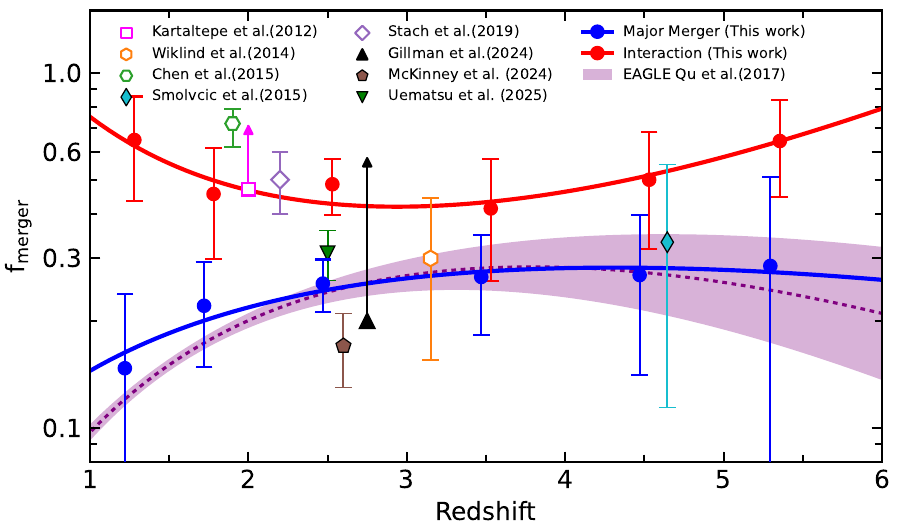} 
\caption{ {\bf Merger Fraction as a Function of Redshift for Our SMGs.} The blue and red points represent the fractions of late-stage major mergers and interactions, respectively, within different redshift bins of our sample. Error bars represent the errors of merger identification and statistical errors. The solid blue and red lines represent the best-fit trends for the evolution of these merger fractions with redshift. The dashed purple lines show the merger fraction evolution for massive (log($M_*/M_\odot$)$>10$) normal galaxies in the EAGLE simulation \citep{Qu2017}. The open magenta squares correspond to the merger fraction of far-infrared selected ULIRGs at $z \sim 2$ \citep{Kartaltepe2012}. The open orange hexagon represents the ongoing mergers among ten SMGs in the GOODS-S field, as detailed by \citet{Wiklind2014}. The open green hexagon denotes ALMA-detected SMGs identified by \citet{Chen2015} in the ECDFS field. The cyan diamond signifies two mergers among six $z>4$ SMGs in the COSMOS field, as reported by \citet{Smolcic2015}. The open purple rhombi indicate the merger fraction of ALMA-detected SMGs in the UKIDSS/UDS field at $z=2.2 \pm 0.1$, according to \citet{Stach2019}. The black triangle depicts the lower limits of the merger fraction for ALMA-detected SMGs from the AS2COSMOS and AS2UDS surveys, as compiled by \citet{Gillman2024}. The brown pentagon represents the merger fraction of SMGs identified by \citet{McKinney2024} in the COSMOS-Web field. The green triangle represents the merger fraction identified by \citet{Uematsu2025} in the COSMOS field. Unfilled markers represent mergers identified from HST/F160W imaging, while filled markers correspond to those identified through JWST and ALMA imaging.
}

\label{fig:Fig04}
\end{figure*}

 In our analysis of the merger fraction, we initially identified galaxies with distinct merger signatures, such as double nuclei, multicomponent structures, or companion galaxies within 15\,kpc exhibiting tidal disturbances. These features indicate that the galaxies are in the late stages of major mergers. JR and FL  independently identified major mergers for this SMG sample. If an SMG was identified as a major merging galaxy by both individuals, it was marked as a `reliable' major merger. If an SMG was considered a merger by only one of us, it was marked as a `tentative' major merger. We identified 49 SMGs as `reliable' major mergers and an additional 10 SMGs as `tentative' major mergers. This classification establishes a major merger fraction of $24.3\pm 3.7\%$ in our sample. This result is consistent with recent identifications of the major merger fraction in the COSMOS and UDS fields \citep{McKinney2024, Gillman2024, Uematsu2025} using JWST data. It is important to note that visual identification is a reliable method for detecting galaxies in the late stages of major mergers. However, we acknowledge that this approach may overlook galaxies in the early stages of merging or those that have already completed the process. Therefore, the merger fraction we report can be considered a conservative estimate.

Furthermore, we identified potential galaxy pairs. For SMGs with companion galaxies within 30\,kpc, we first cross-matched the COSMOS2020 catalog and CANDELS catalogs. If the companion galaxies were cross-matched and had spectroscopic or photometric redshifts, we used the galaxy pair selection method to identify mergers. Specifically, if both the SMG and the companion galaxy have spectroscopic redshifts, their relative velocity $\lvert \Delta V \rvert < 500$ km/s is considered a galaxy pair. When one has a spectroscopic redshift and the other has a photometric redshift, $\lvert \Delta z \rvert$ less than the photometric redshift error $\sigma_z$ is considered a galaxy pair. When both galaxies have photometric redshifts, $\lvert \Delta z \rvert < \sqrt{\sigma^{2}_{1} + \sigma^{2}_{2}}$ is considered a galaxy pair. An SMG is considered a merger and marked as `reliable' if it has at least one companion galaxy within 30 kpc that meets the aforementioned criteria. If an SMG has companion galaxies within 30\,kpc but no previous redshift measurements, it is marked as `tentative'. We also identified galaxies with disturbed morphologies and potential minor mergers based on the presence of distinct clumps at the peripheries of SMGs or irregular dwarf galaxies affected by tidal forces within a 30\,kpc radius. The identification method is similar to that for major mergers. Our sample includes 84 interactions marked as `reliable', and 47 interactions marked as `tentative'. This suggests that $48.4 \pm 11.1\%$ of SMGs may be involved in interactions.

Figure \ref{fig:Fig04} illustrates the evolution of the merger fraction with redshift, accompanied by a compilation of existing data on SMGs or ULIRGs. The figure shows that the fraction of late-stage major mergers is consistent with the predictions of the EAGLE simulation. Our major merger fraction shows a significant increase from 14.7$\pm9.1$\% 
at $z = 1$ to 26.6$\pm 8.4$\% at $z = 3$, and after which it stabilizes and remains approximately constant across the redshift range $3 < z < 6$. We find that the interaction fraction remains relatively stable across the range $2 < z < 5$.

We provide the best-fit relationships between the merger fraction ($f_{\rm mer}$) and redshift for two types of mergers. The merger galaxy fraction as a function of redshift is modeled according to the form presented by \citet{Conselice2009}, which is widely adopted in studies examining the evolution of merger fractions or merger rates with redshift.
The derived $f_{\rm mer} - z$ relations are as follows: for late-stage major mergers (represented by the blue solid line in Figure \ref{fig:Fig04}),  $f_{\rm mer} = (0.08 \pm 0.02)(1+z)^{(1.90 \pm 0.71)} \exp{[-(0.36 \pm 0.17)(1+z)]}$; for interactions (represented by the red solid line in Figure \ref{fig:Fig04}), $f_{\rm mer} = (1.35 \pm 0.31)(1+z)^{(-3.15 \pm 0.69)} \exp{[(0.80 \pm 0.17)(1+z)]}$.  

We emphasize that employing a constant merger fraction offers statistically robust constraints for elucidating the role of mergers in the evolution of SMGs. The evolutionary trends presented in this work should be interpreted as qualitative indicators of merger fraction evolution, rather than definitive quantitative relationships. As demonstrated in Figure \ref{fig:Fig04}, each data point is associated with considerable uncertainties, predominantly stemming from the restricted sample size and systematic limitations inherent in merger identification through visual morphological classification. This observational constraint implies that the current data quality does not warrant the implementation of more sophisticated evolutionary parameterizations, as such approaches would likely introduce substantial risks of overfitting.

\begin{figure*}[ht!]
\centering
\includegraphics[width=0.97\textwidth] {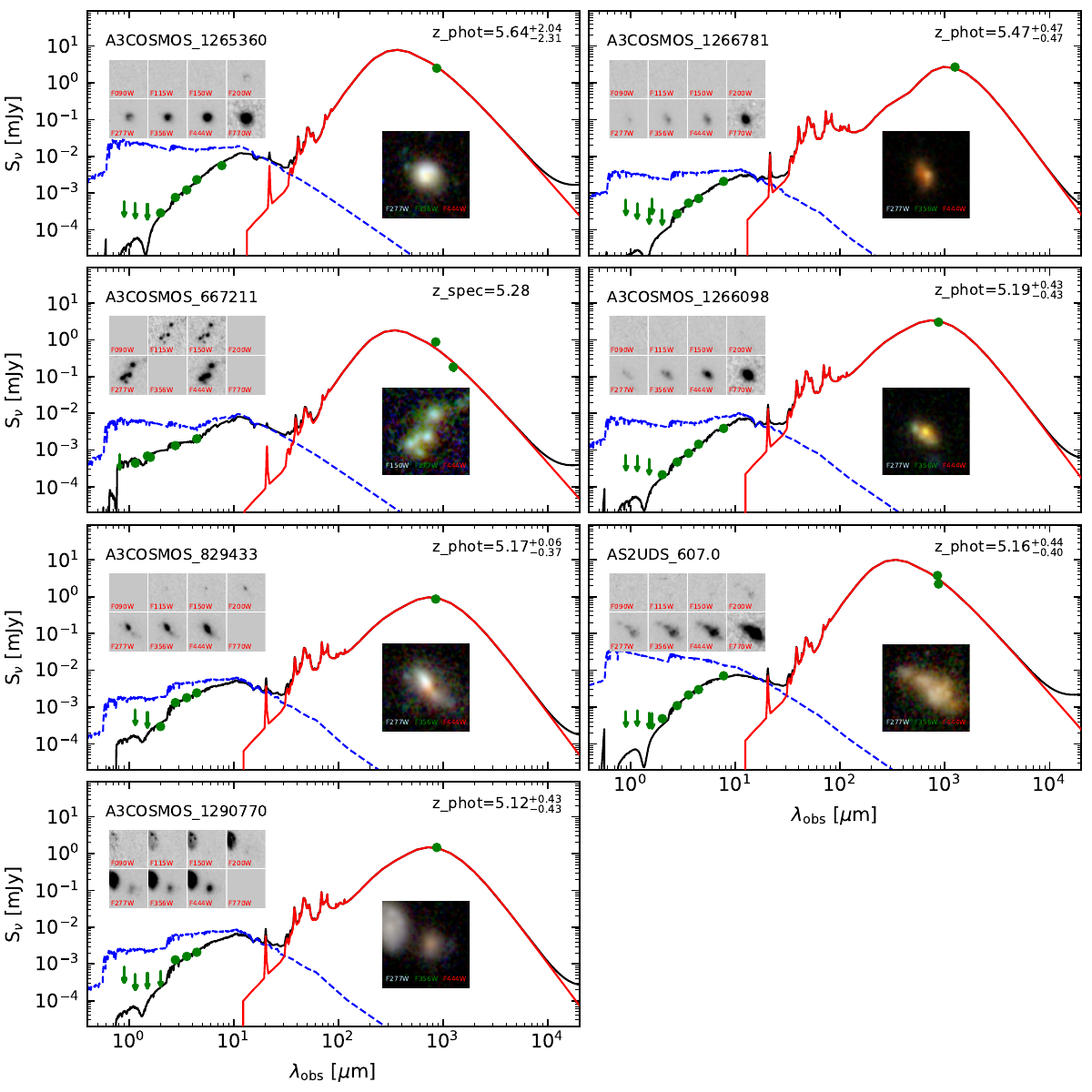} 
\caption{{\bf Multi-band JWST/NIRCam and MIRI Images and Best-fit SEDs of 7 SMGs with Redshifts $z>5$.} The ID and redshift of each SMG are indicated in the upper left corner of the bottom panels. Where available, the effective radius and S\'{e}rsic index $n$ for each band are presented in the upper left corner of the respective images. In each image, the physical size corresponding to the side length is 15\,kpc. }
\label{fig:Fig05}
\end{figure*}

\subsection{Morphology of $z>5$ SMGs}

 We conducted a detailed morphological analysis of seven SMGs with redshifts $z > 5$ in our sample. Figure \ref{fig:Fig05} displays their best-fit Spectral Energy Distributions (SEDs) and multi-band images. Among these seven SMGs, only A3COSMOS\_667211 exhibits multiple components, indicating that it is in a late stage of a major merger. A3COSMOS\_829433 and AS2UDS\_607.0 display structures resembling tidal tails, suggesting that they may have undergone major mergers. A3COSMOS\_1266098 has blue clumps in its outskirts, which may indicate a minor merger, with the outer clumps representing recently accreted dwarf galaxies with high star formation rates and low dust content. A3COSMOS\_1266781 shows a disturbed morphology. In addition, the remaining two, A3COSMOS\_1265360 and A3COSMOS\_1290770, display undisturbed disk-like structures. Thus, among the seven SMGs with redshifts $z > 5$, most show signs of mergers or interactions, implying that the majority of SMGs in the early universe were likely formed through mergers or interactions.

\section{Discussions} \label{sec:Discussion}

\subsection{Impact of Cosmic Dimming effect on Merger Identification} \label{sec:Cosmic Dimming}

The cosmic dimming effect, which follows the form $(1+z)^{-4}$, significantly impacts the detection and classification of galaxies at higher redshifts. For late-stage mergers in our sample, characterized by multi-component structures and strong tidal features, the cosmic dimming effect has a relatively minor impact on their identification. This is because these galaxies have higher surface brightness and more pronounced structural features that are less affected by the dimming effect. However, for minor mergers, the cosmic dimming effect can lead to an underestimation of their fractions as redshifts increase. For galaxy pairs, our SMG sample has a median stellar mass of $\log(M_*/M_\odot) \sim 10^{11}$. Even considering that the companion galaxies have a mass one order of magnitude lower ($\log(M_*/M_\odot) \sim 10^{10}$), the depth of the JWST multi-band imaging is sufficient to detect these galaxy pairs without significantly affecting their estimation. Regarding disturbed morphologies at $z>4$, only two galaxies in our sample were identified as having disturbed morphologies, contributing only 10\% to the total in this redshift range. Even considering a potential overestimation of this 10\%, it does not affect the increasing trend of interactions in the $4<z<6$ range. Additionally, considering the underestimation of minor mergers at higher redshifts, the increasing trend of the interaction fraction in the $4<z<6$ range is even more pronounced.

In conclusion, the cosmic dimming effect and detection limits have a limited impact on our estimates of the major merger fraction. Although there is an underestimation of the interaction fraction at higher redshifts, it does not affect the overall increasing trend. This suggests that our results are robust and provide a reliable understanding of the evolution of merging and interacting galaxies across the redshift range studied.

\begin{figure}[t!]
\centering
\includegraphics[width=0.95\columnwidth] {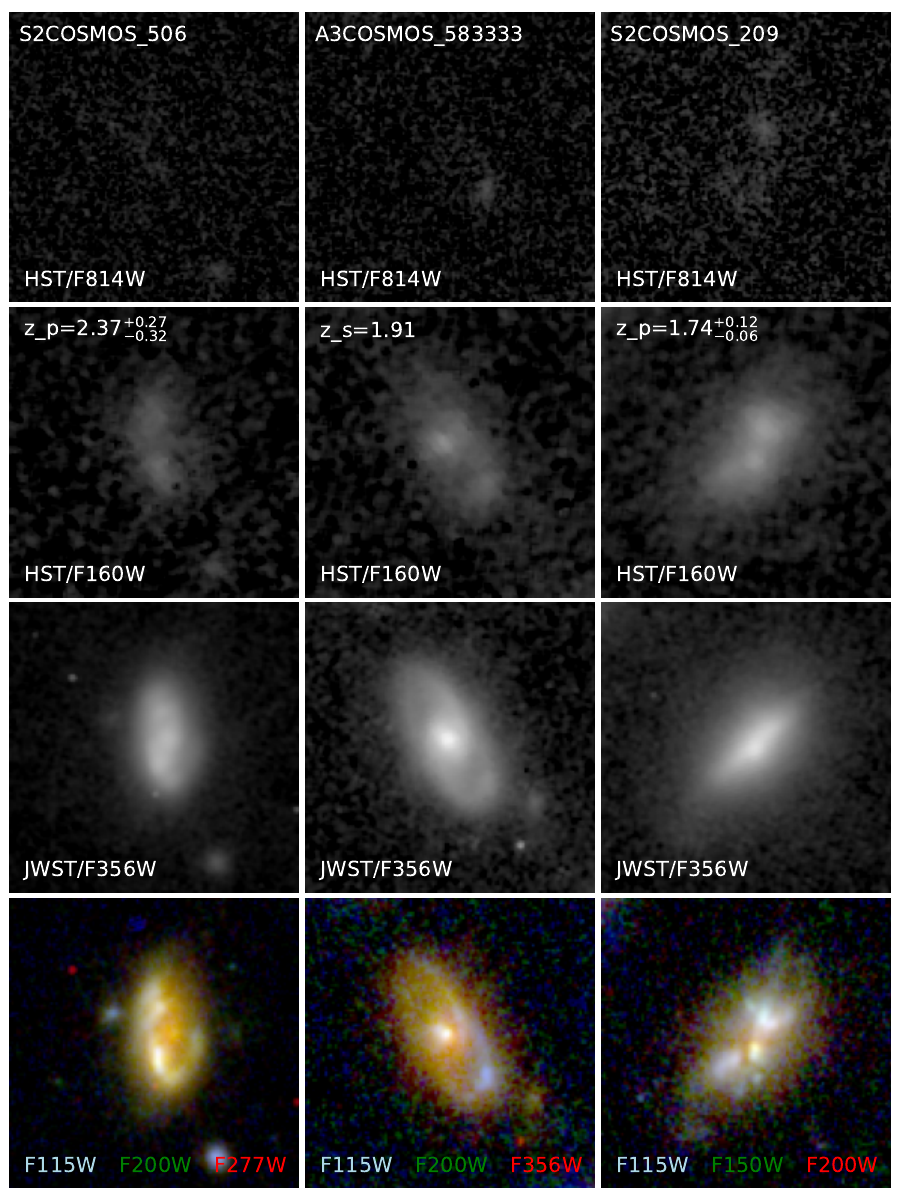} 
\caption{{\bf The HST/F814W and F160W images, as well as longer-wavelength JWST/NIRCam images and pseudo-color images of six typical disk-dominated SMGs at $z\sim2$.}The ID and redshift of each galaxy are indicated in the upper left corner of the bottom panels.  In each image, the physical size of the side length is 30\,kpc. }
\label{fig:Fig06}
\end{figure}

\subsection{Compare the $R_e$ with previous works} \label{sec:Re_evolution}

As illustrated in Figure \ref{fig:Fig02}, the effective radii ($R_e$) of our SMGs at redshifts $z\sim2$ are generally consistent with most previous research on the sizes of SMGs, such as the findings by \citet{Boogaard2023} and \citet{Gillman2023, Gillman2024}. However, as redshift increases, some previous studies, particularly the research by \citet[][hereafter G24]{Gillman2024}, begin to deviate from our measurement results.

 We find that the size-redshift evolution relationship in G24 lies between our rest-frame V and rest-frame NIR-bands. We attribute this primarily to the fact that G24's $R_e$ is around 1 $\mu$m, while our V-band is close to 0.5 $\mu$m and our NIR is close to 1.6 $\mu$m. In addition, even after accounting for a 0.2\,dex underestimation in our stellar mass, G24's median stellar mass is still 0.1\,dex higher than our sample. In fact, after considering both the band and stellar mass, the size-mass relation is consistent with our results. A comparison of our size-mass relation with the results of G24 is provided in Appendix \ref{Size-Mass relation}.

Additionally, we compared the effective radii of SMGs in the rest-frame V-band with those of the general star-forming galaxies (SFGs) recently observed by the JWST. We found that within the redshift range of $1 < z < 5$, the $R_e$ of SMGs in the rest-frame V-band is approximately 1--2\,kpc larger than that of SFGs. This is likely because SMGs typically possess a greater stellar mass and more extensive star formation activity than normal SFGs, and therefore, they have larger effective radii. In the rest-frame near-infrared band, the evolution of our SMG size ($\sim$1.6 $\mu$m) with redshift is consistent with the size ($\sim$ 1.5 $\mu$m) evolution of massive SFGs provided by \citet{Martorano2024} in the redshift range of $1 < z < 2.5$.

\subsection{Previous overestimate of merger fraction at $z\sim2$}

We observe that the merger fractions of SMGs or ULIRGs at redshift $z \sim 2$, as previously determined from HST WFC3/F160W images \citep{Kartaltepe2012, Chen2015, Stach2019}, reach upper limits of our "interaction" fraction. We hypothesize that at redshift $z \sim 2$, the HST/F160W band roughly corresponds to the rest-frame V-band, and the merger-like morphological features exhibited by a substantial number of galaxies may be substructures within the galaxies. These substructures could be misidentified as mergers, resulting in an overestimation of the merger fraction.

 To evaluate the potential overestimation of the merger fraction based solely on HST observational images, we analyzed 58 sources within our sample of SMGs in the redshift range of $1.5 < z < 2.5$ that have HST/F160W observations. We identified 22$\pm$2 major mergers using the HST/F160W images, which account for approximately a major merger fraction of 37.9$\pm$3.4\%. However, by employing multi-band images from JWST, we identified only 13$\pm$1 major mergers, representing a fraction of 22.4$\pm$5.7\%. This suggests that the merger fraction obtained previously using HST may overestimate the true major merger fraction by 69.2\%. Although a considerable fraction of the ``interactions" category may not be mergers in our identification, the fraction obtained from HST/F160W images is 55.2$\pm$5.2\%. In contrast, the result obtained from JWST images is 46.6$\pm$10.8\%, indicating an overestimation of nearly 18.5\%. Figure \ref{fig:Fig06} presents the HST/F814W and F160W images, as well as longer-wavelength JWST/NIRCam images and pseudo-color images of three typical disk-dominated SMGs at $z\sim2$.

\subsection{Formation Mechanisms of SMGs} \label{sec:formation}

 Our study conducted a detailed investigation of the merger fraction of SMGs across a broad redshift range of $z = 1 - 6$, with particular focus on the early cosmic epoch at redshift $z > 5$, when the age of the universe was less than two billion years. During this period, the impact of late-stage major merger events on SMG formation might have been less significant.

Upon comparison with previous studies, we find that our major merger fraction aligns well with the latest merger fractions obtained using JWST/NIRCam confirmations, which are approximately 20\% \citep{Gillman2024, McKinney2024, Uematsu2025}. This fraction is also consistent with the major merger fraction of normal galaxies provided by the EAGLE simulations \citep{Qu2017}, suggesting that major mergers may not be the dominant mechanism for the starburst phase in SMGs.

In addition to our main findings, we observe that the fraction of interacting galaxies within SMGs remains relatively constant in the redshift range $2 < z < 5$. This period coincides with a rapid increase in the cosmic star formation rate density \citep{Madau2014}. The overall star-forming activity in galaxies is intense, allowing a greater number of normal massive disk galaxies to maintain high levels of star formation activity. This activity heats the dust, leading to the formation of submillimeter galaxies (SMGs). In contrast, at higher redshifts ($z> 5$) or lower redshifts ($z < 2$), interactions are necessary to sustain high levels of star formation activity in massive galaxies, which in turn heats the dust. Particularly at lower redshifts, massive galaxies have largely shut down star formation and can only briefly maintain high star formation activity through the merger of a gas-rich galaxy (wet merger).

We find that a substantial fraction of disk SMGs exhibit blue clumps and arcs at cosmic noon. Some SMGs exhibit PSF-like morphologies, implying that they likely have AGN activity. Therefore, we posit that the formation mechanisms of SMGs are primarily consistent with those that sustain high levels of star-forming activity. The intense star formation activity in SMGs can be triggered by instabilities and fragmentation within the gas disk, as well as by mergers and interactions. Additionally, AGN activity contributes to a small fraction of SMGs. However, mergers and interactions not only trigger star formation but also AGN activity. For instance, \citet{Uematsu2025} observed that approximately 47\% of AGNs in SMGs are triggered by major mergers. Consequently, the fraction of SMG formation mechanisms attributed to AGN activity is likely to be half of what is observed in the AGN fraction within SMGs.

\section{Summary}

 In this study, we investigate the evolution of size and merger fraction for a sample of 222 SMGs with dust luminosity of $12 < \log({\rm L}_{\rm IR}/L_\odot) < 13$.  These SMGs were identified through ALMA and JCMT, and their properties were studied using the advanced imaging capabilities of JWST/NIRCam and MIRI across a broad redshift range of $1 < z \lesssim 6$. Our sample has a median redshift of $z = 2.61^{+0.89}_{-0.82}$ and a median stellar mass of $\log(M_*/{\rm M}_\odot) = 10.97^{+0.29}_{-0.33}$. Our key findings are as follows:

\begin{enumerate}

    \item We observe a significant decrease in the effective radius ($R_{\rm e}$) of SMGs with increasing redshift in both the rest-frame V-band and NIR band. The $R_{\rm e}$ of SMGs follows $R_{\rm e} \propto (1 + z)^{-0.87 \pm 0.08}$ in the rest-frame V-band and $R_{\rm e} \propto (1 + z)^{-0.88 \pm 0.11}$ in the NIR band. The NIR-band size evolution of our SMGs resembles that of massive star-forming galaxies at lower redshifts ($1 < z < 2.5$).
        
    \item 
Our analysis reveals a major merger fraction of $24.3\pm 3.7$\% and an interaction fraction 
of $48.4 \pm 11.1\%$ within the sample. The major merger fraction exhibits a significant redshift 
dependence, showing a gradual increase at $z<3$ before reaching a plateau at $3 < z < 6$. 
In contrast, the interaction fraction, which encompasses both major and minor mergers, 
potential pairs, and galaxies with disturbed morphologies, maintains a relatively constant value 
across the redshift range of $2 < z < 5$.

    \item Among the seven SMGs with $z > 5$ in our sample, most exhibit merger or interaction signatures, with only a few showing undisturbed disk-like structures. This implies that the majority of SMGs in the early universe were likely formed through mergers or interactions.
    
    \item HST-based studies may overestimate the major merger fraction by a factor of $\sim 1.7$ at $z \sim 2$. Our analysis using multi-band JWST images suggests that the true merger fraction is lower than previously thought, highlighting the importance of high-resolution, multi-wavelength data in accurately identifying merging systems.

\end{enumerate}

These findings provide valuable insights into the formation and evolution of SMGs across cosmic time, emphasizing the complex interplay between mergers, interactions, and internal processes in driving their observed properties. Our results suggest that late-stage major mergers are not the primary formation mechanism for SMGs at $1 < z \lesssim 6$. Instead, other processes such as gas accretion and interactions likely play a significant role in driving the starburst phase in SMGs at different redshifts.

In conclusion, our study provides a comprehensive analysis of the size and merger fraction evolution of SMGs across a broad redshift range. The derived scaling relations and merger fraction trends offer valuable insights into the formation and evolution of these galaxies, highlighting the need for a multi-faceted approach to understanding their complex histories. Our findings underscore the importance of considering a range of physical processes, including but not limited to mergers, in shaping the properties of SMGs throughout cosmic history.

\section{acknowledgements}

The authors gratefully acknowledge the valuable feedback from the reviewers and the editorial team for enhancing the quality of this work. This project is supported by the National Natural Science Foundation of China (NSFC grants No. 12273052, 11733006, 12090040, 12090041, and 12073051) and the science research grants from the China Manned Space Project (No. CMS-CSST-2021-A04). NL acknowledges the support from the Ministry of Science and Technology of China (No. 2020SKA0110100), the science research grants from the China Manned Space Project (No. CMS-CSST-2021-A01), and the CAS Project for Young Scientists in Basic Research (No. YSBR-062). XZZ acknowledges the support from the National Science Foundation of China (Nos. 12233005 and 12073078) and the science research grants from the China Manned Space Project (Nos. CMS-CSST-2021-A02, CMS-CSST-2021-A04, and CMS-CSST-2021-A07).  FXA acknowledges the support from the National Science Foundation of China (No.12303016). This paper makes use of the following JWST data:  Programs 1727 (PIs Jeyhan Kartaltepe and Caitlin Casey), 1810 (PI Sirio Belli), 1837 (PI James Dunlop), 1840 (PIs Javier Alvarez-Marquez and Takuya Hashimoto),  2514 (PIs Christina Williams and Pascal Oesch) ,and 3990 (PIs Takahiro Morishita, Charlotte Mason, Tommaso Treu, and Michele Trenti).

\bibliography{SMG}{}
\bibliographystyle{aasjournal}

\appendix

\section{The photometry redshifts}

\begin{figure}[h!]
\centering
\includegraphics[width=1.0 \columnwidth] {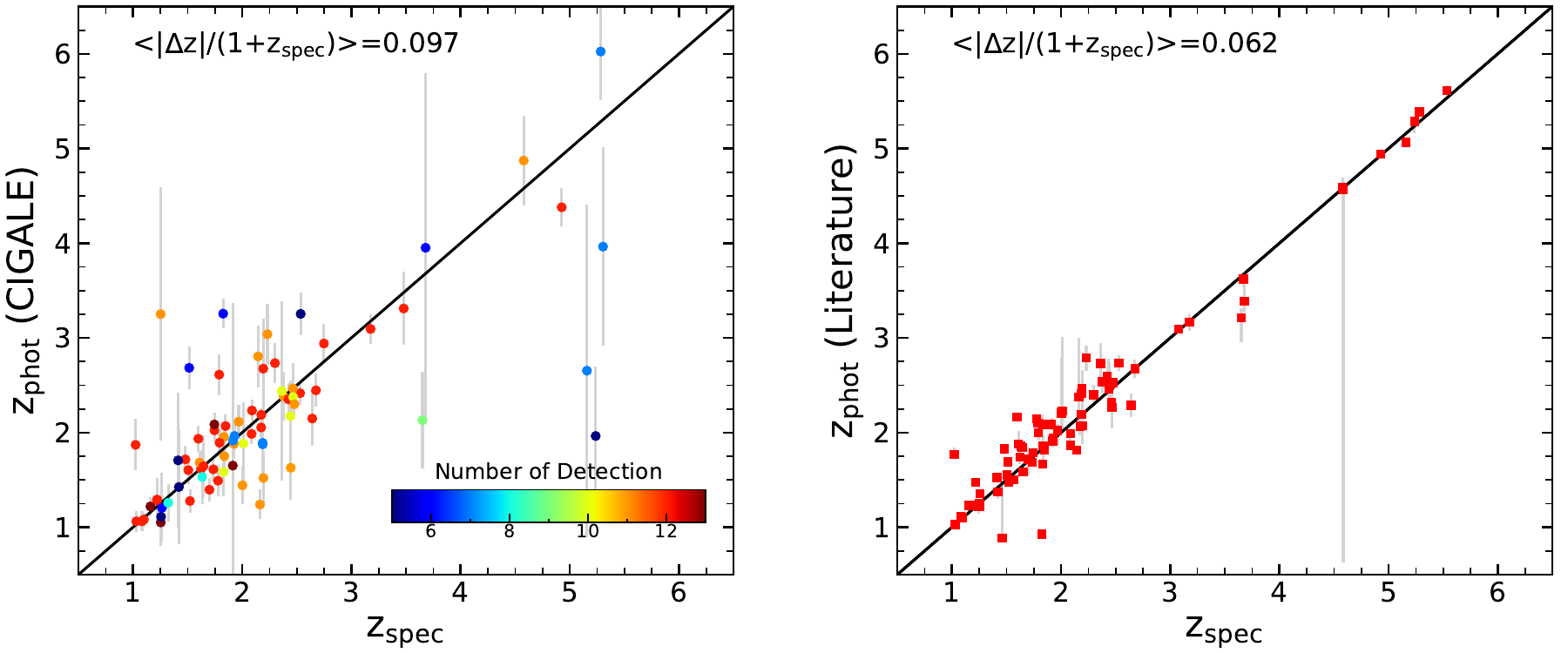} 
\caption{ Left: comparison between spectroscopic redshift and photometric redshift estimated by \texttt{CIGALE}. Right: comparison between  $z_{\rm spec}$ and $z_{\rm phot}$ in literature. 
}
\label{fig:Fig07}
\end{figure}

Figure \ref{fig:Fig07} shows the comparison of the $z_{\rm phot}$ obtained using \texttt{CIGALE} and the best $z_{\rm phot}$ obtained by \citet{Ren2023} using previous data with $z_{\rm spec}$.

\section{Size-Mass relation} \label{Size-Mass relation}

\begin{figure}[h!]
\centering
\includegraphics[width=1.0 \columnwidth] {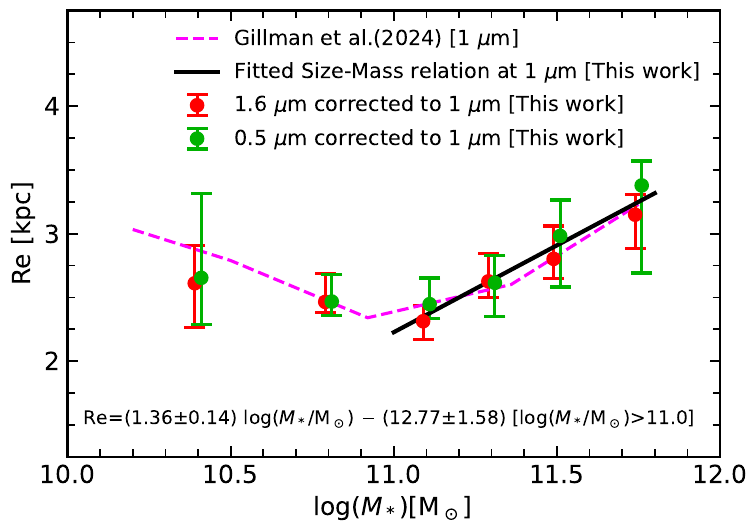} 
\caption{ 
{\bf The restframe 1 $\mu$m size versus the stellar mass of SMGs}. The magenta dashed line indicates the size-mass relation at 1 µm as given by  \citet{Gillman2024} . The green and red points represent the median size of our sample corrected to 1 $\mu$ from rest-frame V and rest-frame NIR-band sizes in each stellar mass bin, respectively. The black solid line denotes the size-mass relation fitted using our data points.
}
\label{fig:Fig08}
\end{figure}

We utilized the size evolution relationship with rest-frame wavelength provided by \citet{Gillman2024} to correct the $R_{\rm e}$ of our SMG sample to rest-frame 1\,$\mu$m. Subsequently, accounting for an underestimation of approximately 0.2\,dex in our stellar mass, we presented the size evolution with stellar mass in Figure \ref{fig:Fig08}. In the log($M_*$/M$_\odot$)$ > 10.5$ regime, our size-mass relation aligns with the one given by \citet{Gillman2024} . We fitted the size-mass relation for our sample at log($M_*$/M$_\odot$)$ > 11$, yielding R$_{\rm e}$ =($ 1.35\pm 0.14 $) log($M_*$/M$_\odot$) - (12.61$\pm $1.64).

\section{ The images of our SMG sample.} 

\begin{figure*}[ht!]
\centering
\includegraphics[width=1\textwidth] {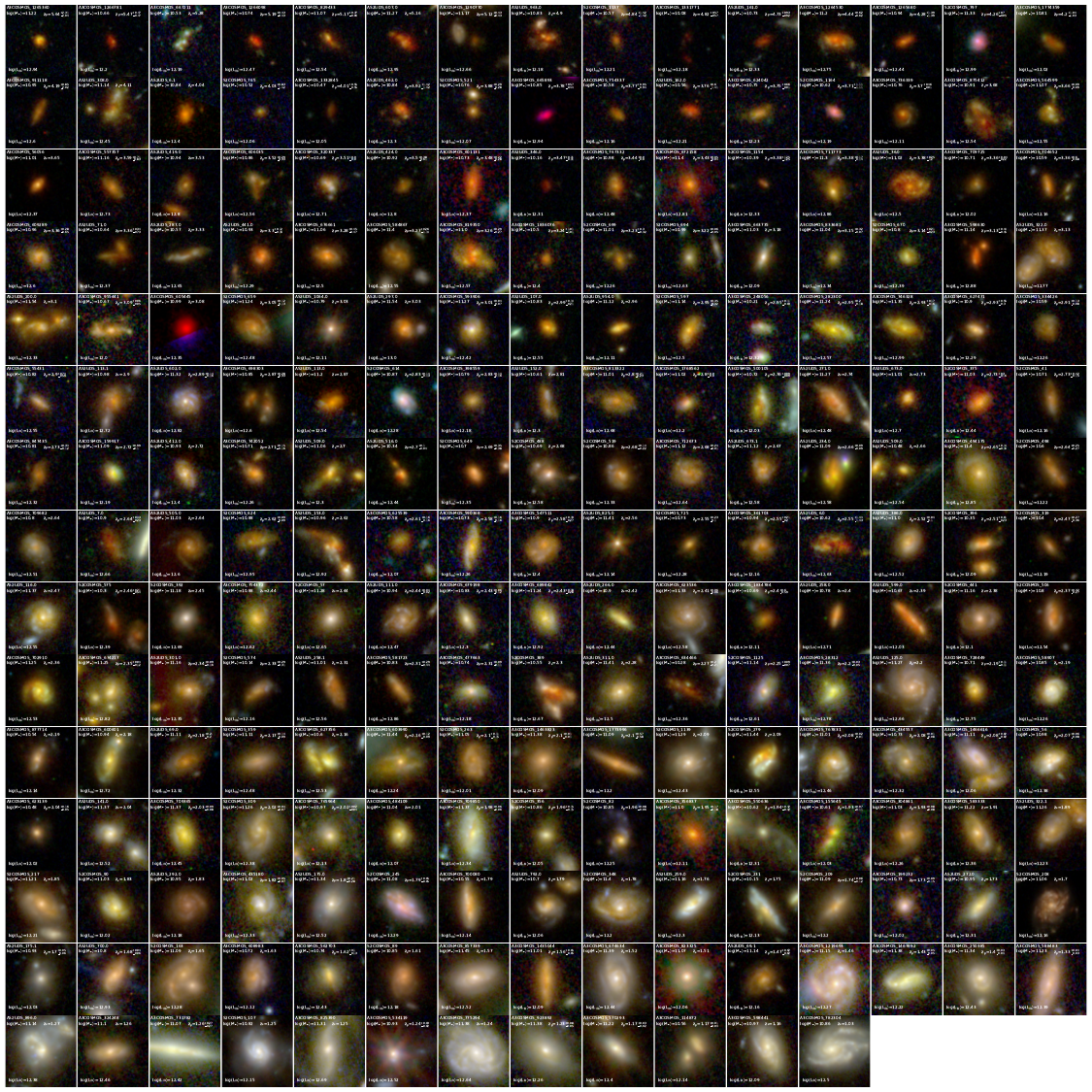} 
\caption{{\bf Pseudo-color images of the identified 222 SMGs.} 
The ID, stellar mass, redshift, and infrared luminosity of each SMG are indicated 
at the top of its image. The images are arranged in redshift descending order.  In each image, the physical size of the side length is 30\,kpc. 
}
\label{fig:Fig09}
\end{figure*}

\end{document}